\documentclass[aps,pre,twocolumn,superscriptaddress,showpacs]{revtex4}
\usepackage{amsfonts,amssymb,amsmath,latexsym,epsfig,times} 
\begin{document}

[Phys. Rev. E {\bf 68}, 056307 (2003)]

\title{Reactive dynamics of inertial particles in nonhyperbolic chaotic flows}

\author{Adilson E.  Motter}
\email{motter@mpipks-dresden.mpg.de}
\affiliation{Max Planck Institute for the Physics of Complex Systems,
N\"othnitzer Strasse 38, 01187 Dresden, Germany}

\author{Ying-Cheng Lai}
\affiliation{Departments of Mathematics, Electrical Engineering, and Physics,
Arizona State University, Tempe, Arizona 85287, USA}

\author{Celso Grebogi}
\affiliation{Instituto de F\'{\i}sica, Universidade de S\~{a}o Paulo,
Caixa Postal 66318, 05315-970 S\~{a}o Paulo, Brazil}

\date{\today}

\begin{abstract}

Anomalous kinetics of infective (e.g., autocatalytic) reactions in open,
nonhyperbolic chaotic flows are important for many applications in biological,
chemical, and environmental sciences.  We present a scaling theory for the
singular enhancement of the production caused by the universal, underlying
fractal patterns.  The key dynamical invariant quantities are the {\it effective
fractal dimension} and {\it effective escape rate}, which are primarily
determined by the hyperbolic components of the underlying dynamical invariant
sets.  The theory is general as it includes
all previously studied hyperbolic reactive dynamics as a special case.
We introduce a class of dissipative embedding maps for numerical verification.

\end{abstract}


\maketitle

Many chemical and biological processes in fluids are characterized by a
filamental distribution of active particles along fractal invariant sets of the
advection chaotic dynamics.  These fractal structures act as dynamical catalysts
for the reaction, which is relevant for a variety of environmental processes in
open flows, such as ozone depletion in the atmosphere \cite{SLLE:1996} and
population dynamics of plankton in the oceans \cite{A:1998}.

The study of active processes in open chaotic flows has attracted a great deal of
interest from the dynamical system community \cite{TKPTG:1998,open_flows}.  Most
of the studies have been performed for time-dependent two-dimensional
incompressible flows, in the limit of weak diffusion.  The flow is nonturbulent,
but the particle dynamics is considered to be chaotic (Lagrangian chaos) and the
active particles to interact with one another without modifying the flow.  The
advection dynamics of such particles can be cast in the context of chaotic
scattering, where incoming tracers spend some time in a mixing (scattering)
region before being scattered along the unstable manifold of the chaotic saddle.
As a result, the products of the reaction concentrate along a fattened-up copy
of the unstable manifold, giving rise to the observed fractal patterns.

Although filamental patterns have been observed in nature,
a clear relation between the observed value of the fractal dimension and the
underlying advection dynamics has been lacking.  For example, in the ``flow past a
cylinder" system previously considered \cite{TKPTG:1998}, the dimension of the
unstable manifold is known to be 2 but the relevant dimension governing
infective and collisional reactions is about 1.6.  This lack of relation,
while not reducing the importance of the previous phenomenological
characterization of filamental distributions of active particles, has led to some skepticism
about the merit of the dynamical system approach to the
problem.  In general, the advection dynamics can be characterized as either
hyperbolic or nonhyperbolic.  In hyperbolic chaotic scattering, all the periodic
orbits are unstable and there are no Kolmogorov-Arnold-Moser (KAM) tori in the
phase space, while the nonhyperbolic counterpart is frequently characterized as
having both chaotic and marginally stable periodic orbits.  Fundamental
assumptions in such works are:  (1) the active particles are massless point-like
tracers and (2) the advection dynamics of these particles is hyperbolic
\cite{inertial}.  However, in realistic situations, the Lagrangian dynamics is
typically nonhyperbolic and the active particles have finite size and inertia.
Indeed, fully hyperbolic systems are quite rare and represent very idealized
situations as the advection dynamics of tracers in fluids is usually constrained
to have a nonhyperbolic character because of no-slip boundary conditions at the
surface of obstacles.  Obstacles are at the same time the origin of Lagrangian
chaos and the origin of nonhyperbolicity.  Even away from obstacles and
boundaries, chaotic motions of tracers typically coexist with regular motions.
In addition, the individual active particles are often too large to be regarded
as noninertial, as is the case for many species of zooplankton in the sea.
Therefore, nonhyperbolic and inertial effects are prevalent in nature
and expected to play an important role in most environmental processes.  A
question of physical importance is then:  what happens to the reactive
dynamics when assumptions (1) and (2) are dropped?

In this article, we present a scaling theory for the reactive
dynamics of inertial particles in nonhyperbolic chaotic flows. The key
concepts in our framework are the {\it effective fractal dimension} and {\it 
effective escape rate}, which are respectively defined as
\begin{eqnarray} 
\label{d_eff}
D_{eff}(\varepsilon )&=&-\frac{d\ln N(\varepsilon)}{d \ln
\varepsilon},\\ \kappa_{eff}(\varepsilon )&=& -\frac{d\ln R(n)}{d
n},
\label{e_eff}
\end{eqnarray}
where $N(\varepsilon)$ is the number of $\varepsilon$-squares needed
to cover the relevant fractal set, and $R(n)$ is the fraction of
particles that takes more than $n=n(\varepsilon )$ steps to escape from
the mixing region (see below).  As a representative application of these
concepts, we show, for autocatalytic reactions of the form $A + B
\rightarrow 2B$, that the area covered by $B$-particles in the steady state
obeys the following scaling law:
\begin{equation} \label{scaling}
{\cal A}_B\sim\left[\frac{\sigma}{e^{\mu_{eff}/(2-D_{eff})}-1}
\right]^{2-D_{eff}},
\end{equation}
where $\mu_{eff}\equiv (\kappa_{eff} + \tilde{\kappa})\tau$, $\tau$
is the time interval between successive reactions (time lag), $\sigma$
is the reaction range, and $\tilde{\kappa}$ is the contraction rate
due to dissipation. For nonhyperbolic flows in two dimensions, $D_{eff}<2$ and
$\kappa_{eff}>0$ are non-trivial functions of the scale $\varepsilon$,
which can often be regarded as constants over a wide interval, even
though $D=\lim_{\varepsilon\rightarrow 0}D_{eff}(\varepsilon)=2$ and
$\kappa=\lim_{\varepsilon\rightarrow 0}\kappa_{eff}(\varepsilon)=0$. 
We find, surprisingly, that
$D_{eff}$ and $\kappa_{eff}$ are significantly different from $D$ and
$\kappa$, respectively, not only for noninertial but also for inertial
particles, even though the advection dynamics of the latter is hyperbolic,
meaning that {\it scars} of the nonhyperbolic conservative dynamics
are observable in the hyperbolic dynamics of slightly dissipative
systems ($\tilde{\kappa}\ll 1$). The previous relations for
noninertial particles in hyperbolic fluids appear as a particular
case of our results.

The nature of the chaotic scattering arising in the context of particle
advection in incompressible fluids may change
fundamentally as the mass and size of the particles are increased from
zero. Physically, this happens because of the detachment of the
particle motion from the local fluid motion. For spherical
particles of finite size, the particle velocity ${\bf v}\equiv d{\bf
x}/dt$ is typically different from the (time dependent) fluid
velocity ${\bf u}={\bf u}({\bf x},t)$ and, in first order, is governed
by the equation \cite{MR:1984}
\begin{equation}
\frac{d{\bf v}}{dt} - \alpha \frac{d{\bf u}}{dt} = - a\;({\bf v} - {\bf u}).
\label{fluid}
\end{equation}
The parameters are
$\alpha = 3\rho_f/(\rho_f+2\rho_p)$
and
$a=\frac{2}{3}\alpha/St$,
where $\rho_f$ and $\rho_p$ are the densities of fluid and particle,
respectively, and $St$ is the Stokes number, which goes to zero in the
limit of point-like particles \cite{approx}. For neutrally buoyant particles, the mass
ratio parameter is $\alpha = 1$, while for aerosols and bubbles, we have
$\alpha < 1$ and $\alpha > 1$, respectively. The inertia
parameter $a$ determines the contraction rate or dissipation in the
{\it phase space} $({\bf x},{\bf v})$, which for incompressible flows
can be shown to be $-2a$. In the limit $a\rightarrow \infty$, the
dynamics is projected on a surface defined by ${\bf v} = {\bf u}$,
which corresponds to the advection dynamics of point particles. The
{\it configuration-space} projection of the particle motion is
strongly influenced by $\alpha$ \cite{BTT:2002}. For small inertia
(large $a$), $\boldsymbol{\nabla}\cdot {\bf v}\approx
a^{-1}(\alpha-1)\boldsymbol{\nabla}\cdot \boldsymbol{(}({\bf
v}\cdot\boldsymbol{\nabla}){\bf u}\boldsymbol{)}
=a^{-1}(\alpha-1)(s^2-\omega^2)$, where $s$ and $\omega$ are
proportional to the strain rate and vorticity of the fluid
\cite{M:1987}, respectively. The behaviors of bubbles and aerosols are 
then qualitatively different. For instance, along a closed orbit,
aerosols are pushed outward, while bubbles are pushed inward.
We first consider bubbles, whose configuration-space dynamics 
is dissipative when the vorticity overcomes the strain rate.

Dynamically, the inertial effects are effectively those due to
dissipation, so that the transition to finite inertia is equivalent to
a transition from open Hamiltonian to dissipative
dynamics. It has been recently shown \cite{ML:2002} that,
while hyperbolic dynamics is robust, nonhyperbolic
chaotic scattering typically undergoes a metamorphosis in the presence of
arbitrarily small amount of dissipation. For nonhyperbolic scattering in
open Hamiltonian systems, particles can spend a long time in the neighborhood
of KAM tori, resulting in an algebraic decay for the survival
probability of particles in the scattering region. As a consequence, the fractal
dimension of the invariant manifolds is the phase-space dimension
\cite{LFO:1991}. This should be contrasted with the
hyperbolic case, whose decay is exponential and fractal dimension is
typically smaller. The dissipation, however, may convert
marginally stable periodic orbits of the KAM islands into
attractors. The survival probability then becomes exponential, the
dimension of the chaotic saddle becomes fractional, and the overall
dynamics of the scattering process becomes hyperbolic.

To understand the meanings of $D_{eff}$ and $\kappa_{eff}$ for fractal sets
arising in the transition from Hamiltonian nonhyperbolic to weakly dissipative 
chaotic scattering, we consider a Cantor set, which is   
constructed in the interval $[0,1]$ according to the rule that 
in the $n$th time step, a fraction $\Delta_n= \gamma/(\beta +n) +\delta$
is removed from the middle of each one of the $N=2^{n-1}$ remaining
subintervals, where $\beta$, $\gamma$, and $\delta$ are constants.
The conservative case corresponds to $\delta=0$, which is
characterized by an algebraic decay with $n$ of the total length
remaining, given by $R(n)\sim n^{-\gamma}$ for $n\gg \beta$, and by a
unity fractal dimension for the invariant set, $D=1$. The removed
fraction $\Delta_n$ decreases at each time step and, as a result, a
systematic change of scales is induced, resulting in a non-self-similar
invariant set that becomes denser as we go to smaller scales. 
The relevant consequence is that the box-counting
dimension converges slowly to 1, leading to a scale-dependent
effective fractal dimension, $D_{eff}\approx
1-\gamma/\ln\varepsilon^{-1}$ for small $\varepsilon$.  Similarly, the
effective escape rate behaves as $\kappa_{eff}\approx \gamma/n \approx
\gamma\ln 2/\ln \varepsilon^{-1}$, where $n=n(\varepsilon)$ is defined
as the number of iterations needed to make the length of each
remaining subinterval smaller than $\varepsilon$.

The limiting dynamics changes drastically and acquires properties of
hyperbolic dynamics when a small amount of dissipation is allowed,
which is modeled by $0<\delta \ll \gamma/\beta$. In particular, the
total length of the remaining intervals decays exponentially,
$R(n)\sim (1- \delta)^n$ for $n\gg \gamma/\delta -\beta$, and the
dimension of the invariant set becomes smaller than one, namely,
$D=\ln 2\boldsymbol{/}\ln{[2/(1-\delta)]}$. At finite time, however, the transition
from the conservative to the dissipative case is much smoother. For
$\ln\varepsilon^{-1}\gg\beta$, the effective fractal dimension and the
effective escape rate are $D_{eff}(\varepsilon )\approx
\ln{2}\boldsymbol{/}\ln{[2/(1-\delta)]} - \gamma'/\ln{\varepsilon^{-1}}$ and
$\kappa_{eff}(\varepsilon )\approx \ln (1-\delta)^{-1} + \gamma'
\ln{[2/(1-\delta)]}\boldsymbol{/}\ln{\varepsilon^{-1}}$, respectively,
where $\gamma'\equiv \gamma/(1-\delta)$. The key feature is that
unrealistically small scales are required to resolve the limiting
values of the fractal dimension and the escape rate, rendering them
physically irrelevant. For instance, to obtain $D_{eff} > 0.95$, scales
$\varepsilon < 10^{-20}$ may be required.  Thus, the physically important
characteristics of the fractal set are the effective dimension and
escape rate.

We now present a physical theory for the scaling law (\ref{scaling}), valid
for autocatalytic reactions in two-dimensional time-periodic flows.
Consider the area covered by $B$-particles in the open part of the
flow (i.e., region where particles eventually escape to infinity) and,
to be specific, that the time lag $\tau$ is integer multiple of the
flow's period. After a sufficiently long time from the onset of the
reaction, the reagent $B$ is distributed along stripes of
approximately uniform width, mimicking the unstable manifold. The
average width $\epsilon$ of the stripes changes aperiodically over
time until the steady state is reached, when it undertakes the
periodicity $\tau$ of the reaction. We assume that the reaction is
sufficiently close to the steady state so that $D_{eff}(\epsilon)$ and
$\kappa_{eff}(\epsilon)$ can be considered constant over time. This
condition is not so restrictive because for many systems $D_{eff}$ and
$\kappa_{eff}$ are essentially constant over several decades (see
below). Therefore, for scales larger than $\epsilon$, the area covered
by $B$-particles can be regarded as a fractal characterized by
dimension $D_{eff}(\epsilon)$ and escape rate
$\kappa_{eff}(\epsilon)$.

Let $\epsilon^{(n-1)}(\tau)$ and $\epsilon^{(n)}(0)$ denote the
average widths of the stripes right before and right after the $n$th
reaction \cite{TKPTG:1998}, respectively. Between successive reactions,
the stripes shrink due to escape and dissipation as follows:
$\epsilon^{(n)}(\tau)=\epsilon^{(n)}(0) e^{-h_{eff}\;\tau}$,
where $h_{eff}=(\kappa_{eff}+\tilde{\kappa})/(2-D_{eff})$ plays the
role of an {\it effective} (contracting) Lyapunov exponent, while $\tilde{\kappa}$
accounts for the nonconservative contribution. When the reaction occurs, the
widening due to the reaction is proportional to the reaction range:
$\epsilon^{(n+1)}(0)-\epsilon^{(n)}(\tau)\propto\sigma$.
The area covered by $B$-particles right before the $(n+1)$th reaction
${\cal A}_B^{(n)}$ satisfies
${\cal A}_B^{(n)}\propto\boldsymbol{(}\epsilon^{(n)}(\tau)\boldsymbol{)}^{2-D_{eff}}$.
These relations can then be combined to
yield the following recursive relation for the area:
${\cal A}_B^{(n+1)}=e^{-\mu_{eff}}\left[ ({\cal
A}_B^{(n)})^{1/(2-D_{eff})}+c\;\sigma \right]^{2-D_{eff}}$,
where $\mu_{eff} = (\kappa_{eff} +\tilde{\kappa})\tau$ and $c$ is a
constant geometric factor.  From the condition ${\cal
A}_B^{(n+1)}={\cal A}_B^{(n)}$ follows our main scaling (\ref{scaling}) 
for the area ${\cal A}_B$ in the steady state \cite{cont}.
This scaling holds for both noninertial and inertial
particles, regardless of whether the flow is hyperbolic or nonhyperbolic.
The hyperbolic case with inertial particles is studied in
Ref. \cite{TNMGT:PNAS}. The scaling  (\ref{scaling}) represents a
further step toward generality
since it is also valid for nonhyperbolic flows.

To make possible a numerical verification of the scaling law (\ref{scaling}), 
it is necessary at present to use discrete-time maps. To construct a class of
maps that captures all essential features of continuous-time chaotic flows, we
note: (1) the fluid dynamics, determined by $d{\bf
x}/dt={\bf u}$, is {\it embedded} in the particle's advection equation
and is recovered in the limit $a\rightarrow \infty$; (2) the
phase-space contraction is determined by $a$ (irrespective of
$\alpha$); (3) for small inertia, the configuration-space contraction
is proportional to $a^{-1}(\alpha-1)$.
For an area preserving map ${\bf x}_{n+1}={\bf M}({\bf x}_n)$,
representing the dynamics of a time-periodic incompressible fluid, a
possible choice for the corresponding {\it embedding map} representing
the {\it inertial} particle dynamics is
${\bf x}_{n+2} - {\bf M}({\bf x}_{n+1}) =
e^{-a}\boldsymbol{(}\alpha{\bf x}_{n+1}-{\bf M}({\bf x}_n)\boldsymbol{)}$,
where the factors involving $a$ and $\alpha$ are naturally imposed by
the particle dynamics \cite{c10}. This can be written as
\begin{eqnarray}
\label{embed2x}
{\bf x}_{n+1} &=& {\bf M}({\bf x}_{n}) + \boldsymbol{\delta}_n,\\
\boldsymbol{\delta}_{n+1}&=&e^{-a}\boldsymbol{ (}\alpha{\bf x}_{n+1}-{\bf M}({\bf x}_n)\boldsymbol{ )},
\label{embed2d}
\end{eqnarray}
where ${\bf x}$ and $\boldsymbol{\delta}$ can be interpreted as the
configuration-space coordinates and the detachment from the fluid
velocity, respectively, so that (${\bf x}$, $\boldsymbol{\delta}$)
represents the phase-space coordinates. 
This class of embedding maps
can be a paradigm to address many problems in inertial advection dynamics
as it captures the essential properties of
Eq.~(\ref{fluid}).  In particular, it is uniformly dissipative, with
phase-space contraction rate equal to $e^{-2a}$; the noninertial
dynamics ${\bf x}_{n+1} = {\bf M}({\bf x}_{n})$ is recovered in the
limit $a\rightarrow \infty$; and the configuration-space contraction rate
is proportional to $e^{-a}(\alpha-1)$ for $e^{-a}(\alpha-1)\ll 1$, in
agreement with the distinct behavior expected for aerosols and
bubbles.  Therefore, for finite $a$, a rich higher dimensional dynamics
with $\alpha$-dependent ${\bf x}$-space projection is expected. Next
we consider such a dynamics for both $\alpha > 1$ and $\alpha < 1$.

To simulate the flow, we consider a two-dimensional area-preserving map that has
a pronounced nonhyperbolic character \cite{LFO:1991}:  $(x,y)\rightarrow
[\lambda(x-w^2/4),\lambda^{-1}(y+w^2)]$, where $w\equiv x+y/4$ and $\lambda>1$
is the bifurcation parameter.  The dynamics is nonhyperbolic for
$\lambda\lesssim 6.5$.  For $\lambda=4$, for example, there is a major KAM
island in the $xy$-space, as shown in Fig.~\ref{fig1}(a).  Also, from
Fig.~\ref{fig1}(a), one can see tangencies between the stable and unstable
manifolds in the neighborhood of the KAM island, which is a signature of
nonhyperbolicity.  It is well established that, within the nonhyperbolic region,
the dimension of the invariant manifolds is $D=2$ and the escape rate is
$\kappa=0$ \cite{LFO:1991,ML:2002}.  When this map is embedded in
Eqs.~(\ref{embed2x}-\ref{embed2d}), for $\alpha > 1$, the $xy$-projection of the
resulting 4-dimensional map is dissipative in the mixing region (KAM islands and
neighborhood).  In this regime, the dissipation stabilizes marginally stable
periodic orbits in the KAM islands of the conservative map, converting the KAM
islands and neighborhood into the corresponding basin of attraction of the newly
created attractors, as shown in Fig.~\ref{fig1}(b).  The basin itself extends
around the mixing region mimicking the stable manifold of the conservative
dynamics.  As a result, the tangencies between the invariant manifolds
apparently disappear, suggesting that the advection dynamics of bubble particles
is {\it hyperbolic}.   For $\alpha < 1$, on the other hand, the
configuration-space projection expands in the mixing region and almost all the
orbits eventually escape to infinity.  However, for small inertia and $\alpha$
close to 1, particles in the regions corresponding to KAM islands of the
conservative dynamics and neighborhood are {\it almost trapped} in the sense
that the time it takes to escape is much larger in these
regions than outside
them.  These regions are neglected in our analysis of the open part of the flow,
as shown in Fig.~\ref{fig1}(c), because filamental structures cannot be resolved
inside them.

Numerical simulation of the autocatalytic reaction is performed by dividing the
mixing region with a grid where the size of the cells represents the reaction
range $\sigma$.  Particles are placed in the center of the cells.  When a
reaction takes place in a cell occupied by a $B$-particle, all the cells
adjacent to it are {\it infected} with $B$-particles \cite{c4}.  We assume that
$A$ is the background material and that the reaction takes place simultaneously
for all the particles at time intervals $\tau$ \cite{c3}.  If we start with a small seed
of $B$-particles near the stable manifold, after a transient time a steady state
is reached where $B$-particles are accumulated along a fattened-up copy of the
unstable manifold, as shown in Fig.~\ref{fig1}(d) for massless point particles,
in Fig.~\ref{fig1}(e) for bubbles, and in Fig.~\ref{fig1}(f) for aerosols.  In
the computation, particles $A$ and $B$ are set to have the same mass ratio and
inertia parameters.  To compute the effective fractal dimension $D_{eff}$ of
the unstable manifold, we use the uncertainty algorithm \cite{MGOY:1985} applied
to the first order approximation of the inverse map.  The effective dimension
turns out to be constant over many orders of magnitude of variations in
$\varepsilon\;$ and it is approximately the same for both noninertial and
slightly inertial bubble particles ($D_{eff}= 1.73$ for $\varepsilon >
10^{-15}$), while it is somewhat smaller for slightly inertial aerosol particles
($D_{eff}= 1.68$ for $\varepsilon > 10^{-15}$), as shown in Fig.~\ref{fig2}(a).
Strong evidence of the scaling law (\ref{scaling}) is presented in
Fig.~\ref{fig2}(b) for two different values of the time lag $\tau$,
where the scaling exponent is consistent with
$D_{eff}= 1.73$ for noninertial and bubble particles, and with $D_{eff}= 1.68$ for
aerosol particles.  We see that even though the area ${\cal A}_B$ changes with
the inertial properties of the particles, the {\it scaling} of ${\cal A}_B$
remains essentially the same for bubbles, as expected from our
Cantor-set model.

It is instructive to compare this result with the reaction-free dynamics.  The
dimension of nonhyerbolic invariant sets can be argued to be integer by mean of
a zoom-in technique, where a fast numerical convergence is achieved by focusing
on the densest parts of the fractal \cite{LFO:1991}.  The reaction, however, has
a {\it global} character, as it takes place along the unstable manifold around
the entire mixing region.  This makes the convergence of the relevant effective
dimension extremely slow, and that is why the effective dimension is apparently
constant.

In summary, we have shown that the dynamical system approach to the reactive
dynamics in imperfectly mixed flows also applies to realistic situations where
nonhyperbolic and inertial effects are relevant.  The rate equations of reactive
processes are primarily governed by finite-time dynamics and as such change
smoothly in the {\it noninertial} $\rightarrow$ {\it inertial} transition, which
is in sharp contrast with the metamorphosis undergone by the long-term and
asymptotic dynamics of reaction-free particles.  We have focused on
autocatalytic reactions, but our results are expected to hold whenever the
reaction front mimics the underlying unstable manifold and the activity takes
place along the boundary of a fattened-up fractal.  Examples of this kind of
process include infective reactions (e.g., combustion \cite{KMSSKN:2003}) and
collisional reactions in general (e.g., $A+B\rightarrow 2C$, where an unlimited
amount of material $A$ is present and material $B$ is continuously injected in
the vicinity of the stable manifold \cite{TKPTG:1998}).  Finally, we observe
that our analysis does not rely on the existence of a well-defined fractal set
in the advection dynamics.  The results remain valid as long as effective values
for the fractal dimension and escape rate can be properly defined and are
approximately constant over the relevant interval of observation.  This is
important for environmental processes, whose underlying dynamics is only
partially understood \cite{TNMGT:PNAS}.

AEM and YCL were supported by AFOSR under Grant No. F49620-03-1-0290. CG was supported by Fapesp and CNPq.
AEM thanks Tamas T\'el for illuminating discussions.

\begin{figure}[pt]
\begin{center}
\caption
{ For $\lambda = 4$, (a) KAM island (light grey), and stable (grey) and unstable (black) 
  manifolds for $a=\infty$.  (b) Fixed point attractor (black
  dot), basin of attraction (grey), and unstable manifold (black), for
  $\alpha=1.05$ and $a=1$. (c) Stable (grey) and unstable (black) manifolds 
  for
  $\alpha=0.95$ and $a=1$, outside the region covered by the almost trapped orbits (light grey). Particles
  are launched with initial velocity matching the fluid velocity 
  ($\boldsymbol{\delta}_{0}={\bf 0}$).
  (d-f) Corresponding area covered by $B$-particles in the ``open" part of the flow right before
        the reaction, for $\tau=5$ and $\sigma=5.10^{-3}$. }
\label{fig1}
\end{center}
\end{figure}

\begin{figure}[ph]
\begin{center}
\caption
{For $\lambda = 4$, (a) Effective dimension of the unstable 
     manifold as computed from the uncertainty method,
     where $f(\varepsilon)$ is the fraction of $\varepsilon$-uncertain points
     in the line $x=0$, $0<y<0.1$ of the time reversed dynamics.
 (b) Scaling of the relative area ${\cal A}_B$ covered by $B$-particles
     [in the region shown in Figs.~\ref{fig1}(d)-(f), 
     right before the reaction] as a function of the 
     reaction range $\sigma$ for two choices of the time lag $\tau$. 
     In both plots, stars correspond to 
     noninertial particles ($a=\infty$), circles to 
     bubble particles with $\alpha=1.05$ and $a=1$, and plus signs to aerosol
     particles with $\alpha=0.95$ and $a=1$. The aerosol data in (a) are
     shifted vertically downward for clarity.}
\label{fig2}
\end{center}
\end{figure}


\begin{references}
\bibitem{cont}
A similar scaling law can be derived
in the continuous-time limit $\tau\rightarrow 0$.
\end{references}
\end{document}